\documentclass[%
 reprint,
superscriptaddress,
amsmath,amssymb,
aps,
pra,
]{revtex4-1}

\usepackage{graphicx}
\usepackage{dcolumn}
\usepackage{bm}

\begin{document}

\preprint{APS/123-QED}

\title{Changing the speed of optical coherence in free space}

\author{Murat Yessenov}
\author{Ayman F. Abouraddy}
\email{Corresponding author: raddy@creol.ucf.edu}
\affiliation{CREOL, The College of Optics \& Photonics, University of Central Florida, Orlando, Florida 32186, USA}




\begin{abstract}
It is typically assumed that the fluctuations associated with a stationary broadband incoherent field propagate in free space at the speed of light in vacuum $c$. Here we introduce the concept of `coherence group velocity', which -- in analogy to the group velocity of coherent pulses -- is the speed of the peak of the coherence function. We confirm experimentally that incorporating a judicious spatio-temporal spectral structure into a field allows tuning its coherence group velocity in free space. Utilizing light from a super-luminescent diode, we interferometrically measure the group delay encountered by the cross-correlation of a structured field synthesized from this source with the unstructured diode field. By tracking the propagation of this cross-correlation function, we measure coherence group velocities in the range from $12c$ to $-6c$. 
\end{abstract}

\maketitle

What is the speed of optical coherence? To put this question on a concrete basis, consider first the case of a pulsed coherent field. The \textit{group velocity} of such a pulse -- the speed of its peak -- can deviate from $c$ in slow-light and fast-light structures \cite{Boyd09Science,Kurgin08Book}, in the case of spatially structured pulses \cite{Giovannini15Science,Bouchard16Optica}, and `space-time' (ST) wave packets in which the spatio-temporal spectrum is structured so that each spatial frequency is assigned to a single wavelength \cite{Donnelly93PRSLA,Saari04PRE,Longhi04OE,Kondakci17NP,Yessenov19PRA,Yessenov19OPN} (see also \cite{FigueroaBook14,SaintMarie17Optica}). In the latter case, unprecedented control over the group velocity of a pulsed beam has been recently demonstrated \cite{Kondakci19NC,Bhaduri19Optica,Yessenov19OE}, including tuning the group velocity in free space from $30c$ to $-4c$. In the case of a \textit{broadband} incoherent stationary field, there are no temporal \textit{intensity} features to be tracked to assess a `coherence group velocity'. However, one may track the propagation of the field coherence function, whose velocity in free space is expected to be $c$ \cite{Mandel95Book}. What happens when stationary incoherent light traverses a slow-light medium or when the spatio-temporal spectral structure associated with ST wave packets is introduced into the field? To the best of our knowledge, the former scenario has not been tested, likely due to the narrow bandwidth of slow-light systems. However, broadband \textit{incoherent} ST fields have been recently synthesized and their non-diffracting behavior confirmed \cite{Yessenov19Optica}.

The question remains whether the coherence group velocity of a ST field -- the speed of the peak of its coherence function -- is $c$ or otherwise. Specifically, consider the scenario where an incoherent ST field and a reference incoherent field (from which the ST field is derived) are made to interfere. After adding a distance $L_{1}$ exceeding the coherence length in the path of the reference field, interference is no longer observed. What is the distance $L_{2}$ that must be placed in the path of the ST field in order to retrieve the interference? Intuitively, one expects that $L_{2}\!=\!L_{1}$. If $L_{2}\!\neq\!L_{1}$ is required to regain interference, then the coherence group velocity associated with the propagation of the underlying temporal fluctuations in the broadband ST field is no longer $c$.

Here we show that the coherence group velocity $v_{\mathrm{c}}$ of a broadband incoherent stationary ST field can indeed differ substantially from $c$ in free space. By introducing a judicious spatio-temporal spectral structure into light from a super-luminescent diode (SLD) and recording the cross-correlation of this field with unstructured light obtained directly from the SLD, we tune $v_{\mathrm{c}}$ in free space to arbitrary values -- whether superluminal, subluminal, or even negative. Because these interference effects are a consequence of the fluctuations underlying the stationary field, our work therefore demonstrates that spatio-temporal structuring of the field can modify the propagation of these fluctuations, including traveling backwards.

\begin{figure}[t!]
\centering
\includegraphics[width=8.6cm]{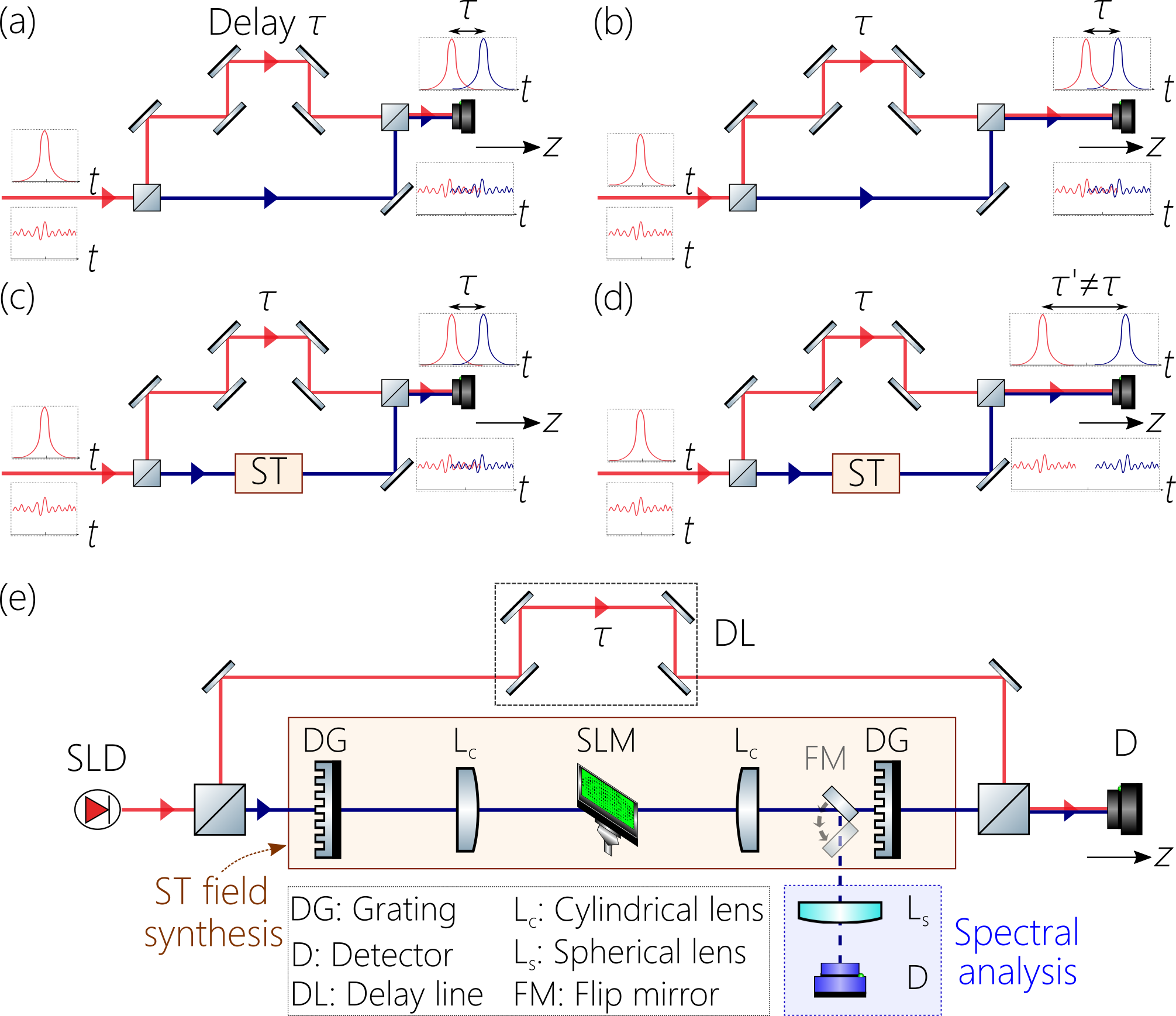}
\caption{\small{Concept of coherence group velocity. (a) A pulse or temporally incoherent field in a MZI with a detector placed at $z\!=\!0$. (b) Same as (a) after moving the detector at the output to $z\!=\!L$. There is no change in the detected interferogram. (c) A pulse or temporally incoherent field enters a MZI with a spatio-temporal synthesis system in one arm. The recorded interferogram with the detector located at $z\!=\!0$ at the output is the cross-correlation between the ST field and the initial field. (d) Same as (c) after moving the detector along $z$. Surprisingly, the interference is lost and the MZI delay must be shifted to retrieve the interference. (e) Schematic of the setup for synthesizing incoherent ST fields starting with a SLD, and for measuring the speed of optical coherence in free space.}}
\label{Fig:Concept}
\end{figure}

We first examine the generic scenario in Fig.~\ref{Fig:Concept}(a) where a pulse \textit{or} temporally incoherent field enters a Mach-Zehnder interferometer (MZI), and sweeping a delay $\tau$ in one arm traces an interferogram corresponding to the field autocorrelation (plus a constant term). The location of the detector after the MZI is \textit{not} relevant, and the interferogram is recorded by sweeping the \textit{same} values of $\tau$ regardless of detector location [Fig.~\ref{Fig:Concept}(b)]. The common path after the MZI does \textit{not} introduce a relative delay because the fields from both arms propagate at the same velocity.

Now consider the scenario in Fig.~\ref{Fig:Concept}(c) where one arm of the MZI includes a system that synthesizes a ST wave packet from a coherent pulse \cite{Kondakci17NP,Kondakci18OE,Bhaduri18OE} or a ST field from a temporally incoherent field \cite{Yessenov19Optica}. Surprisingly, if the detector is displaced along $z$ [Fig.~\ref{Fig:Concept}(d)] when employing a coherent pulsed source, the interference vanishes \cite{Kondakci19NC}. Although the additional path length is \textit{common} to both fields, no interference is observed because of the deviation of the group velocity of the ST wave packet from $c$ \cite{Kondakci19NC,Bhaduri19Optica, Yessenov19OE}. We conjecture that the same effect occurs with temporally incoherent fields because the underlying fluctuations shared between the reference and the synthesized ST field cause them to interfere when overlapping in space and time. This suggests that the \textit{coherence group velocity} $v_{\mathrm{c}}$ of the ST field -- the velocity of the peak of the observed interferogram -- can differ from $c$. We proceed to confirm this conjecture.

Consider a plane-wave broadband temporally incoherent stationary field $E(t)$ from a SLD (or LED after spatial filtering \cite{Yessenov19Optica}), whose coherence function is $G(\tau)\!=\!\langle E(t)E^{*}(t-\tau)\rangle$, where $\langle\cdot\rangle$ represents a statistical average. In the Fourier domain the field is $E(z,t)\!=\!e^{i(k_{\mathrm{o}}z-\omega_{\mathrm{o}}t)}\!\!\int\!d\Omega\widetilde{\psi}(\Omega)e^{-i\Omega(t-z/c)}\!=\!e^{i(k_{\mathrm{o}}z-\omega_{\mathrm{o}}t)}\psi(0,t-\tfrac{z}{c}$); where $\omega_{\mathrm{o}}$ is a fixed temporal frequency, $k_{\mathrm{o}}\!=\!\tfrac{\omega_{\mathrm{o}}}{c}$ is the corresponding free-space wave number, $\Omega\!=\!\omega-\omega_{\mathrm{o}}$, $\psi(0,0,t)$ is the Fourier transform of $\widetilde{\psi}(\Omega)$, $z$ is the axial coordinate, and $E$ is independent of the transverse coordinates. The coherence function is then given by $G(z,\tau)\!=\!\langle E(z,t)E^{*}(z,t-\tau)\rangle\!=\!\!e^{-i\omega_{\mathrm{o}}\tau}\!\int\!d\Omega\widetilde{G}(\Omega)e^{-i\Omega\tau}$, where $\widetilde{G}(\Omega)\!=\!\langle|\widetilde{\psi}(\Omega)|^{2}\rangle$; i.e., the different frequencies are uncorrelated. Except for an additional constant term, the real part of this autocorrelation function $G(z,\tau)$ is the interferogram detected at the output of the MZI in Fig.~\ref{Fig:Concept}(a). Because $G(z,\tau)$ is \textit{independent} of $z$, the \textit{same} interferogram is detected at the output in Fig.~\ref{Fig:Concept}(b) for \textit{any} detector position $z$.

The fundamental distinctive feature of ST fields is that each temporal frequency $\omega$ is associated with a spatial frequency (transverse component of the wave vector) $\pm k_{x}$ \cite{Kondakci19OL}. Here $x$ is a transverse coordinate and we assume that the field is uniform along $y$ ($k_{y}\!=\!0$) for simplicity. The dependence of $k_{x}$ on $\omega$ results from confining the spatio-temporal spectrum to the intersection of the light-cone $k_{x}^{2}\!+\!k_{z}^{2}\!=\!(\tfrac{\omega}{c})^{2}$ with a spectral plane $\Omega/c\!=\!(k_{z}\!-\!k_{\mathrm{o}})\tan{\theta}$ making a spectral tilt angle $\theta$ with respect to the $k_{z}$-axis \cite{Donnelly93PRSLA,Kondakci17NP,Yessenov19PRA}. The relationship between $k_{x}$ and $\omega$ takes the form of a conical section: an ellipse when $0\!<\theta\!<45^{\circ}$ and a hyperbola when $45^{\circ}\!<\!\theta\!<\!135^{\circ}$. The ST field therefore takes the form $E_{\mathrm{ST}}(x,z,t)\!=\!e^{i(k_{\mathrm{o}}z-\omega_{\mathrm{o}}t)}\!\!\int\!d\Omega\widetilde{\psi}(\Omega)e^{ik_{x}(\Omega,\theta)x}e^{-i\Omega(t-\tfrac{z}{c}\cot{\theta})}\!=\!e^{i(k_{\mathrm{o}}z-\omega_{\mathrm{o}}t)}\psi_{\mathrm{ST}}(x,0,t-z/v_{\mathrm{c}})$, where $v_{\mathrm{c}}\!=c\tan{\theta}$, $k_{x}(\Omega,\theta)$ is the appropriate $\theta$-dependent conic section \cite{Yessenov19PRA}, and $\psi_{\mathrm{ST}}(0,0,t)$ is the Fourier transform of $\widetilde{\psi}(\Omega)$. The cross correlation between the delayed reference field $E(x,z,t-\tau)$ and the ST field $E_{\mathrm{ST}}(x,z,t)$, $G_{\mathrm{ST}}(x,z,\tau)\!=\!\langle E_{\mathrm{ST}}(x,z,t)E^{*}(x,z,t-\tau)\rangle$, is
\begin{eqnarray}\label{eq:STCoherence}
G_{\mathrm{ST}}(x,z,\tau)\!\!&=&\!\!e^{-i\omega_{\mathrm{o}}\tau}\!\int\!d\Omega\widetilde{G}(\Omega)e^{ik_{x}(\Omega,\theta)x}e^{-i\Omega(\tau-\Delta\tau)}\nonumber\\
&=&\!G_{\mathrm{ST}}(x,0,\tau-\Delta\tau),
\end{eqnarray}
where $\Delta\tau\!=\!z(\tfrac{1}{c}-\tfrac{1}{v_{\mathrm{c}}})$, and the spatio-temporal profile of $G_{\mathrm{ST}}(x,z,\tau)$ does \textit{not} change with propagation. The interferogram detected at the output of the MZI in Fig.~\ref{Fig:Concept}(c,d) is the real part of Eq.~\ref{eq:STCoherence} plus a constant term. Note that this interferogram includes an additional delay $\Delta\tau$ unrelated to the MZI delay $\tau$ that does not appear in the traditional scenario [Fig.~\ref{Fig:Concept}(a,b)] and results here because of the difference between the coherence group velocity of the reference field $c$ and that of the ST field $v_{\mathrm{c}}$. The location of the detector $z$ therefore surprisingly plays a critical role in determining the interferogram.

In our experiments we make use of light from a fiber-coupled SLD (Inphenix IPSDD0809; power $\sim\!13$~mW, central wavelength $\sim\!800$~nm, bandwidth $\sim\!12$~nm). The arrangement to synthesize a ST field from this incoherent stationary source is the same we previously employed for coherent pulses \cite{Yessenov19OPN}. The system is reminiscent of spectral phase modulation of ultrashort pulses \cite{Weiner09Book} except that the dimension orthogonal to the that of the spread spectrum at a spatial light modulator (SLM, Hamamatsu X10468-02) is exploited to assign a pair of spatial frequencies $\pm k_{x}$ to each wavelength $\lambda$. The functional form of this spatio-temporal spectral association $k_{x}(\lambda,\theta)$ is sculpted such that the axial component of the `group' velocity $v_{\mathrm{c}}\!=\!\tfrac{\partial\omega}{\partial k_{z}}$ takes on arbitrary values: below $c$ (subluminal) when $0^{\circ}\!<\!\theta\!<\!45^{\circ}$, above $c$ (superluminal) when $45^{\circ}\!<\!\theta\!<\!90^{\circ}$, or even negative-$v_{\mathrm{c}}$ when $\theta\!>\!90^{\circ}$ \cite{Salo01JOA,Saari04PRE,Longhi04OE,Yessenov19PRA,Yessenov19OE}. Tuning the spectral tilt angle $\theta$, and thus the coherence group velocity $v_{\mathrm{c}}$, requires only changing the phase imparted by the SLM to the incident spread spectrum. The entire synthesis arrangement is then placed in one arm of a MZI [Fig.~\ref{Fig:Concept}(e)], with an optical delay $\tau$ placed in the reference arm where unfiltered light from the SLD traverses. A CCD camera (TheImagingSource, DMK 27BUP031) is placed at the output to detect the spatially resolved interference fringes.

\begin{figure*}[t!]
\centering
\includegraphics[width=16.8cm]{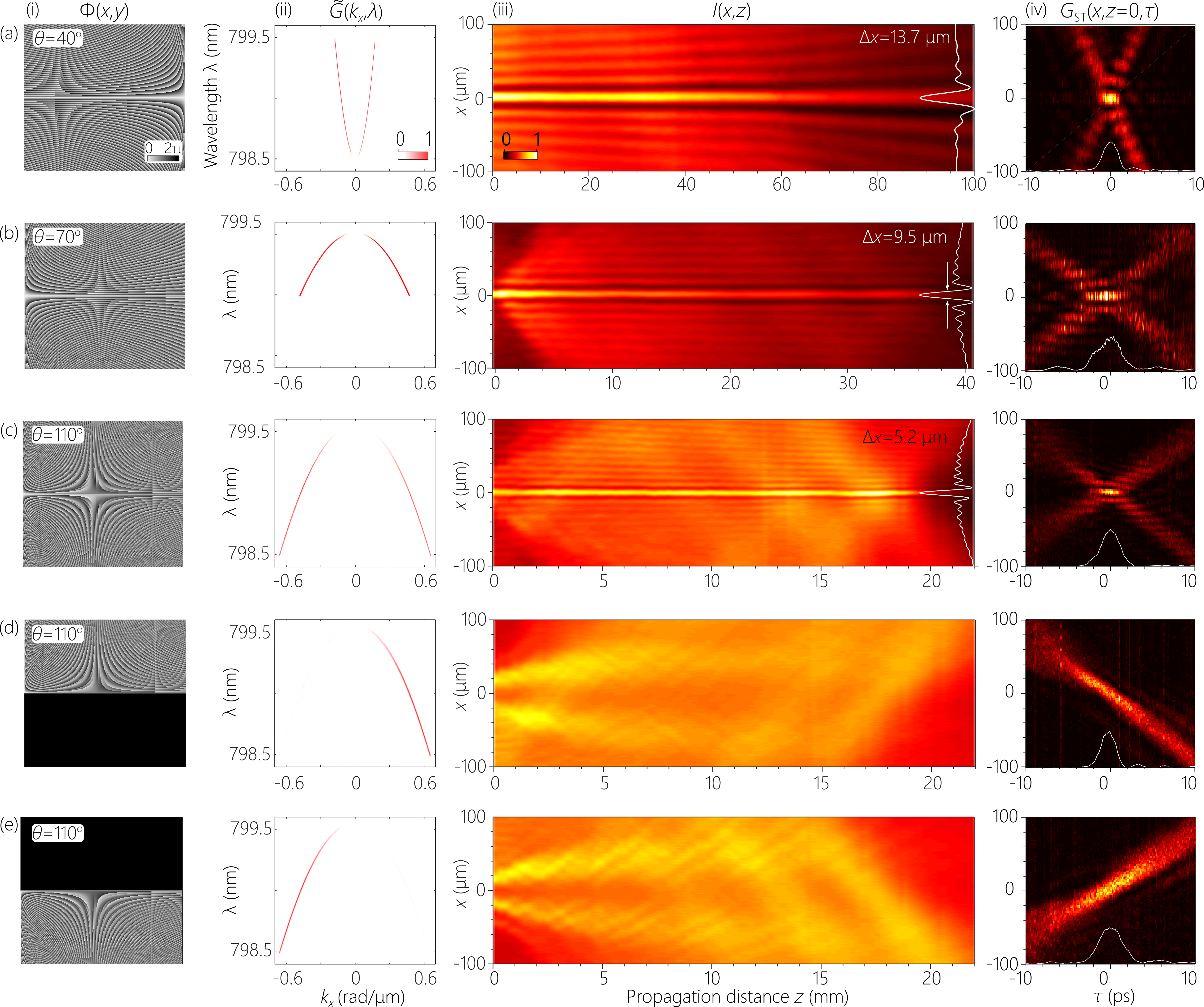}
\caption{Measurements for five ST fields: (a) subluminal with $\theta\!=\!40^{\circ}$ and $v_{\mathrm{c}}\!\approx\!0.84c$, (b) superluminal with $\theta\!=\!70^{\circ}$ and $v_{\mathrm{c}}\!\approx\!2.75c$, (c) negative-$v_{\mathrm{c}}$ with $\theta\!=\!110^{\circ}$ and $v_{\mathrm{c}}\!\approx\!-2.75c$, (d) same as (c) but with only positive-$k_{x}$ values, and (e) same as (c) with only negative-$k_{x}$ values. Column (i) shows the SLM phase distribution, (ii) the measured spatio-temporal spectrum $\widetilde{G}(k_x,\lambda)=\langle|\widetilde{\psi}(\lambda)|^{2}\rangle\delta(k_{x}-k_{x}(\lambda,\theta))$, (iii) the measured time-averaged intensity distribution $I(x,z)$ highlighting the width of the central spatial feature, and (iv) the measured spatio-temporal profile $G_{\mathrm{ST}}(x,z,\tau)$ at $z\!=\!0$. The spectra in (ii) correspond to projections onto the $(k_{x},\tfrac{\omega}{c})$-plane of the intersection of the light-cone $k_{x}^{2}+k_{z}^{2}\!=\!(\tfrac{\omega}{c})^{2}$ with a spectral plane $\omega-\omega_{\mathrm{o}}\!=\!c(k_{z}-k_{\mathrm{o}})\tan{\theta}$ making an angle $\theta$ with the $k_{z}$-axis.}
\label{Fig:TemporalCoherence}
\end{figure*}

We plot in Fig.~\ref{Fig:TemporalCoherence} measurements for five ST fields. Figure~\ref{Fig:TemporalCoherence}(a) shows the results for a subluminal ST wave packet where $\theta\!=\!40^{\circ}$ and $v_{\mathrm{c}}\!\approx\!0.84c$. We provide the 2D phase distribution imparted by the SLM to the incident spread spectrum of the field, the measured spatio-temporal spectrum, the measured propagating time-averaged intensity $I(x,z)$ obtained by scanning the CCD camera along $z$, and the measured spatio-temporal cross-correlation retrieved from the MZI, $G_{\mathrm{ST}}(x,z\!=\!0,\tau)$. The bandwidth is $\sim\!1$~nm, resulting in a coherence time of $\sim\!2$~ps. The associated spatial bandwidth is $\Delta k_{x}\!\sim\!0.18$~rad/$\mu$m, resulting in a spatial feature of width 13.7~$\mu$m \cite{Yessenov19Optica}. We present similar measurements for a superluminal ST wave packet with $\theta\!=\!70^{\circ}$ and $v_{\mathrm{c}}\!\approx\!2.75c$ [Fig.~\ref{Fig:TemporalCoherence}(b)], and a negative-$v_{\mathrm{c}}$ ST wave packet with $\theta\!=\!110^{\circ}$ and $v_{\mathrm{c}}\!\approx\!-2.75c$ [Fig.~\ref{Fig:TemporalCoherence}(c)]. In all three cases, a central peak in the intensity profile is prominent. However, the existence of this spatial feature in the time-averaged intensity $I(x,z)$ is \textit{not} critical for measuring $v_{\mathrm{c}}$. Indeed, as demonstrated in \cite{Yessenov19Optica}, the central spatial feature is a consequence of the mutual coherence between the positive and negative spatial frequencies $\pm k_{x}$, which are in fact derived from the field amplitude for one wavelength from the source. Therefore, when one half of the spatial spectrum is blocked, the intensity distribution is uniform with no spatial features [Fig.~\ref{Fig:TemporalCoherence}(d,e)]. The spatio-temporal coherence measurements nevertheless reveal a tilted cross-correlation function. Whereas tilted pulse fronts are well-known \cite{Fulop10Review,Wong17ACSP2,Kondakci19ACSP}, we believe that this is the first observation of a \textit{tilted optical coherence front}. The coherence group velocity $v_{\mathrm{c}}$ is then estimated by tracking the motion of the time-resolved cross-correlation function $G_{\mathrm{ST}}(x,z,\tau)$.

\begin{figure}[t!]
\centering
\includegraphics[width=8.6cm]{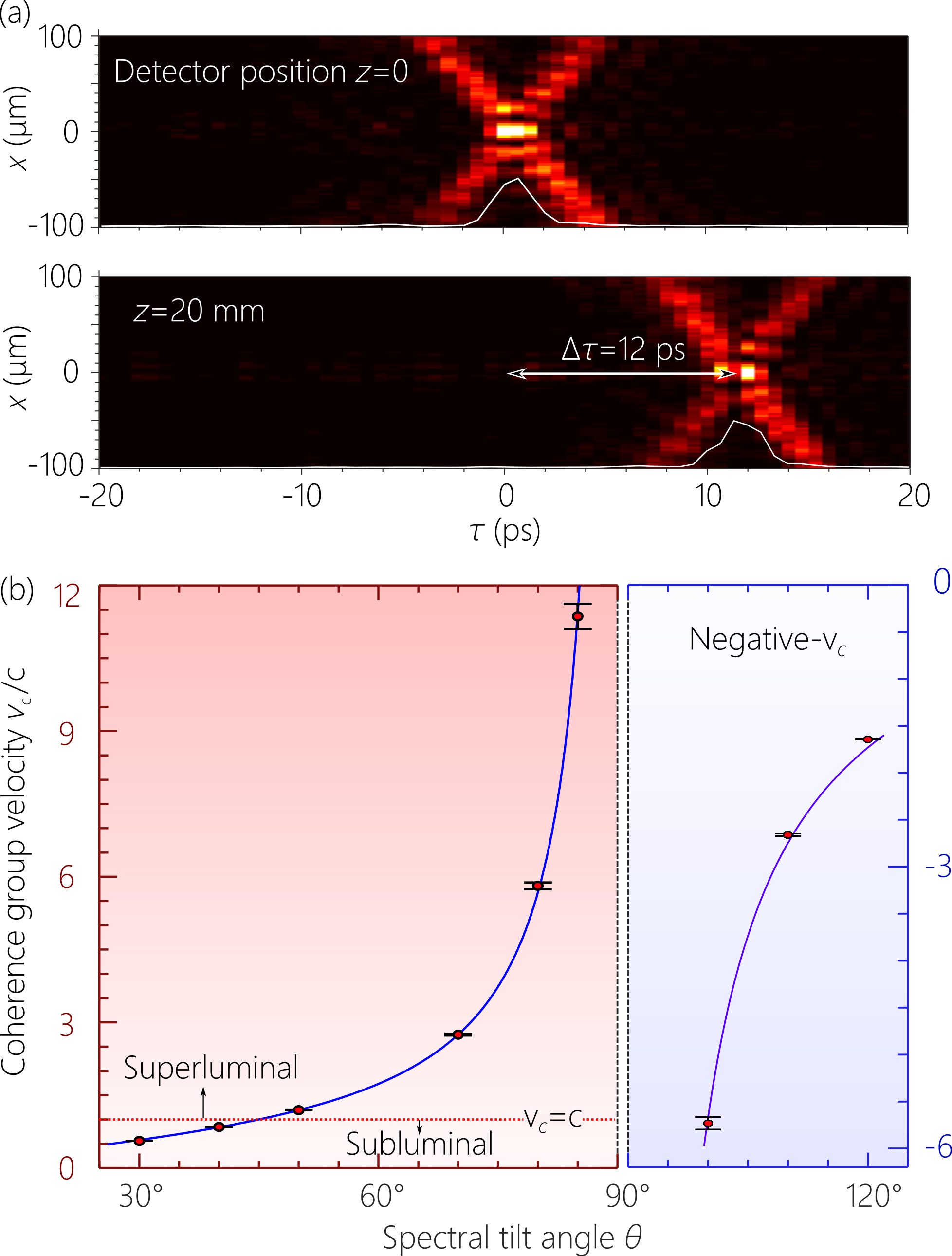}
\caption{(a) Estimating the coherence group velocity $v_{\mathrm{c}}$ when $\theta\!=\!40^{\circ}$ (subluminal with $v_{\mathrm{c}}\!=\!0.84c$ from the propagation of the cross-correlation function $G_{\mathrm{ST}}(x,z,\tau)$. The center of the interferogram has shifted by $\Delta\tau\!\approx\!12$~ps after displacing the detector from $z\!=\!0$ to $z\!=\!20$~mm; i.e., the ST field as expected arrives \textit{later} than the reference unstructured SLD field that propagates at $c$. The white curves are the on-axis values of the cross-correlation function, $G_{\mathrm{ST}}(x\!=\!0,z,\tau)$. (b) Measured $v_{\mathrm{c}}$ with the spectral tilt angle $\theta$. The left panel shows the positive-$v_{\mathrm{c}}$ regime $0^{\circ}\!<\!\theta\!<\!90^{\theta}$ (both subluminal and superluminal), and the right panel the negative-$v_{\mathrm{c}}$ regime $\theta\!>\!90^{\circ}$. The solid curve is the theoretical expectation $v_{\mathrm{c}}\!=\!c\tan{\theta}$.}
\label{Fig:MeasuringSpeed}
\end{figure}

The measurement results shown in Fig.~\ref{Fig:TemporalCoherence} are all obtained with the detector placed at $z\!=\!0$ at the MZI output. To obtain the coherence group velocity $v_{\mathrm{c}}$, we displace the detector to a new position $z$ and re-scan the MZI delay $\tau$. As described above [Fig.~\ref{Fig:Concept}(c,d) and Eq.~\ref{eq:STCoherence}], a surprising feature of the MZI setup here in stark contrast to traditional optical interferometry is the sensitivity of the interferogram to the \textit{absolute} location of the detector in the \textit{common path} because $v_{\mathrm{c}}$ differs from $c$. In Fig.~\ref{Fig:MeasuringSpeed}(a) we plot the measured cross-correlation function for $\theta\!=\!40^{\circ}$ for detector positions $z\!=\!0$  and $z\!=\!20$~mm. Note that the center of the interferogram has shifted by $\Delta\tau\!\approx\!12$~ps with respect to a field propagating at $c$, corresponding to a coherence group velocity of $v_{\mathrm{c}}\!\approx\!0.85c$. By repeating this measurements for different values of $\theta$, we obtain coherence group velocities that are subluminal (to $\approx\!0.5c$), superluminal (to $\approx\!12c$), and negative (to $\approx\!-6c$), with excellent agreement between the measured values of $v_{\mathrm{c}}$ and the theoretical prediction $v_{\mathrm{c}}\!=\!c\tan{\theta}$ as shown in Fig.~\ref{Fig:MeasuringSpeed}(b).

Pioneering work by Saari reported diffraction-free ST fields (X-waves \cite{Saari97PRL} and focus-wave modes \cite{Reivelt00JOSAA,Reivelt02PRE}) using broadband incoherent light, but a coherence group velocity was not assessed. Partially incoherent versions of such fields have been examined theoretically in Refs.~\cite{Friberg91PRA,Saastamoinen09PRA}. Our work is related to the dark and antidark incoherent beams proposed theoretically by Ponomarenko \cite{Ponomarenko07OL}, and which have been recently subject of further experimental \cite{Yessenov19Optica,Zhu19OL} and theoretical \cite{Hyde19JOSAA,Yaalou19OQE} investigations. Here we obtain \textit{time-resolved} cross-correlation measurements that reveal their underlying spatio-temporal structure. Finally, Jedrkiewicz \textit{et al.} \cite{Jedrkiewicz06PRL,Jedrkiewicz07PRA} demonstrated X-shaped coherence functions of light produced by the quadratic nonlinear process of parametric generation resulting from a laser interacting with a second-harmonic crystal -- in contrast to our linear approach with incoherent light from a SLD -- by taking the Fourier transform of the measured spatio-temporal \textit{spectrum}, but without identifying the propagation velocity. Furthermore, we have measured here -- for the first time -- the coherence function in space and time directly without taking the Fourier transform of a spectrum, a procedure that introduces ambiguities because the spectral coherence phase -- associated with dark beams for instance \cite{Ponomarenko07OL,Yessenov19Optica} -- is ignored.  

In conclusion, we have introduced the concept of coherence group velocity for incoherent fields in analogy to the group velocity of coherent pulses. By sculpting the spatio-temporal spectrum of a broadband temporally incoherent stationary field, the coherence group velocity can be readily tuned in free space, which may be useful in optical metrology. Utilizing light from a SLD, we have recorded coherence group velocities in the range from $12c$ to $-6c$, spanning the subluminal, superluminal, and negative-$v_{\mathrm{c}}$ regimes by simply varying the phase distribution imparted to the field spatio-temporal spectrum via a SLM.

\section*{Funding}
US Office of Naval Research (ONR) contracts N00014-17-1-2458 and N00014-19-1-2192.



%

\end{document}